\documentclass[showpacs,prd]{revtex4}
\usepackage{epsfig}
\usepackage{graphicx}%
\usepackage{dcolumn}
\usepackage{amsmath}
\usepackage{latexsym}

\begin{document}

\title{Interactions and non-commutativity in quantum Hall systems}
\author{Frederik G. Scholtz} 
\altaffiliation{fgs@sun.ac.za}
\author{Biswajit Chakraborty} 
\altaffiliation{biswajit@bose.res.in}
\affiliation{Institute of Theoretical Physics, University of
Stellenbosch,\\ Stellenbosch 7600, South Africa}
\affiliation{Satyendra Nath Bose National Centre for Basic Sciences, \\
Block-JD, Sector-III, Salt Lake, Kolkata - 700098, India.}
\author{Sunandan Gangopadhyay}
\altaffiliation{sunandan@bose.res.in}
\affiliation{Satyendra Nath Bose National Centre for Basic Sciences, \\
Block-JD, Sector-III, Salt Lake, Kolkata - 700098, India.}
\author{Jan Govaerts}
\altaffiliation{Jan.Govaerts@fynu.ucl.ac.be}
\affiliation{Institute of Nuclear Physics, Department of Physics, Catholic University of Louvain 2, Chemin du Cyclotron, B-1348 Louvain-la-Neuve, Belgium.}
\date{\today}

\begin{abstract}
We discuss the role that interactions play in the non-commutative structure 
that arises when the relative coordinates of two interacting particles are 
projected onto the lowest Landau level. It is shown that the interactions 
in general renormalize the non-commutative parameter away from the 
non-interacting value $\frac{1}{B}$. The effective non-commutative parameter 
is in general also angular momentum dependent. An heuristic argument, 
based on the non-commutative coordinates, is given to find the filling 
fractions at incompressibilty, which are in general renormalized by the interactions, and the results are consistent with known results in the case of singular magnetic fields.  
\pacs{11.10.Nx} 

\end{abstract}

\maketitle

\section{Introduction}
\label{intro}

The quantum Hall system has recently attracted considerable attention from 
the point of view of non-commutative quantum mechanics and quantum 
field theory \cite{myung}-\cite{jellal} as it is probably the simplest 
physical realization of a non-commutative spatial geometry. Some time ago 
Dunne, Jackiw and Trugenberger \cite{dunne} already observed this 
non-commutativity by noting that in the limit $m\rightarrow 0$ 
the $y$-coordinate is effectively constrained to the momentum canonical 
conjugate to the $x$-coordinate. This result can also be 
obtained \cite{myung1,ouvry1} by keeping the mass fixed and taking 
the limit $B\rightarrow\infty$. An alternative point of view is to keep 
the magnetic field and mass finite, but to project the position coordinates 
onto the lowest (or higher) Landau level. These projected operators 
indeed satisfy the commutation relation ($\hbar=e=m=c=1$)\cite{ouvry2} 
\begin{equation}
\label{commutator}
\left[P_0xP_0,P_0yP_0\right]=\frac{1}{iB}P_0=\frac{1}{iB}
\end{equation}    
where $P_0$ denotes the projector onto the lowest Landau level, which is 
also just the identity operator on the projected subspace, as reflected 
in the last step.

This result allows a simple heuristic understanding of quantum Hall fluids. 
Recall the elementary uncertainty relation (see {\it e.g.} \cite{gaz}) 
for two non-commuting operators $A$ and $B$
\begin{eqnarray}
\label{uncertain}
(\Delta A)^2(\Delta B)^2&\geq& \frac{1}{4}\langle i[A,B]\rangle^2\,,\\
(\Delta A)^2=\langle A^2\rangle-\langle A\rangle^2\,,&&(\Delta B)^2=\langle B^2\rangle-\langle B\rangle^2\,,\nonumber
\end{eqnarray}
where $\langle\cdot\rangle$ denotes the normalized expectation value in 
some state. Using this we note that the non-commutativity of the coordinates 
implies a lower bound to the area a particle in the lowest Landau 
level occupies. This bound follows easily from (\ref{uncertain}) to be 
$\Delta A= 4|\Delta (P_0xP_0)||\Delta (P_0yP_0)|\geq\frac{2}{B}
\equiv \Delta A_0$. This implies that the number of states available in 
a Landau level is $M=\frac{A}{\Delta A_0}=\frac{AB}{2}$. The filling 
fraction is defined in the usual way as $\nu=\frac{N}{M}=\frac{2N}{AB}$, 
with $N$ the number of electrons. For fermions it then follows that at 
maximum filling of $p$ Landau levels the particles must occupy the minimal 
allowed area, {\it i.e.}, $N\Delta A_0=pA$ and $\nu=p$ with $p$ integer. 
As the next available states are in the higher Landau level, separated 
in energy by the cyclotron frequency, one expects that the quantum fluid 
will be incompressible at these values of the filling fraction.
  
In the literature mentioned above the effect that interactions between 
electrons might have on the non-commutativity was not investigated. 
In Ref.\cite{bigatti} an harmonic potential between two interacting 
particles was considered, but there the non-commutativity of the center 
of mass coordinates was investigated, which is again, effectively, 
a Landau problem. In particular the analysis of Ref.\cite{ouvry2} has 
been done in the absence of any interactions between particles. Given 
the conjectured equivalence between a non-commutative $U(1)$ Chern-Simons 
theory \cite{susskind,hellerman} and the composite fermion description for 
the fractional Hall effect, which is an effective non-interacting theory 
for the interacting quantum Hall system, one would like to understand 
the relationship between non-commutativity and interactions better. 
A similar picture arises in the much simpler setting of non-commutative 
quantum Hall systems where it seems as if the fractional quantum Hall 
effect, associated with an interacting quantum Hall system, can effectively 
be described by a non-interacting non-commutative quantum Hall 
system \cite{jellal}, again suggesting an interplay between 
non-commutativity and interactions. Indeed, keeping the philosophy of 
the composite fermion picture in mind, which replaces electrons 
interacting through a short ranged repulsive interaction by 
non-interacting composite fermions moving in a reduced magnetic field, 
one would expect that the interactions must modify the commutation 
relation (\ref{commutator}) as the magnetic field is reduced. A similar 
conclusion was reached from a completely different point of view in 
a recent article of the present authors \cite{scholtz}. Here we want 
to investigate this question in more detail using the approach 
of Ref.\cite{ouvry2}.

To set the scene we consider two interacting particles with the same masses 
and charges moving in a plane with constant magnetic field perpendicular 
to the plane. In the symmetric gauge the Hamiltonian is given 
by ($\hbar=e=m=c=1$)
\begin{eqnarray}
\label{ham1}
H &=&\frac{1}{2}\left(\vec{p}_1-\vec{A}(\vec{x}_1)\right)^2+
\frac{1}{2}\left(\vec{p}_2-\vec{A}(\vec{x}_2)\right)^2+
V(|\vec{x}_1-\vec{x}_2|)\;,\\
A_{i}(\vec{y})&=&-\frac{B}{2}\epsilon_{ij}y^{j},\;B\geq 0\;.\nonumber
\end{eqnarray}
Introducing relative and center of mass coordinates through
\begin{equation}
\vec{R}=\frac{1}{2}(\vec{x}_1+\vec{x}_2),\quad \vec{r}=\vec{x}_1-\vec{x}_2,
\end{equation}
the Hamiltonian reduces to
\begin{eqnarray}
\label{ham2}
H &=&\frac{1}{4}\left(\vec{P}-2\vec{A}(\vec{R})\right)^2+
\left(\vec{p}-\frac{1}{2}\vec{A}(\vec{r})\right)^2+V(|\vec{r}|),\\
\vec{P}&=&(\vec{p}_1+\vec{p}_2),\quad \vec{p}=
\frac{1}{2}\left(\vec{p}_1-\vec{p}_2\right)\;.\nonumber
\end{eqnarray}
This is of course two decoupled problems. The center of mass motion 
corresponds to that of a particle with mass $M=2$ and charge $q=2e=2$ 
moving in a magnetic field $B$, while the relative motion is that of 
a particle with reduced mass $\mu=\frac{m}{2}=\frac{1}{2}$ and 
charge $q=\frac{e}{2}=\frac{1}{2}$ moving in a magnetic field $B$ and 
radial potential $V(|\vec{r}|)$. The cyclotron frequency for both problems 
is $B$. The center of mass motion can clearly be analysed as 
in Ref.\cite{ouvry2}; projection onto the lowest Landau level will lead 
to non-commutative coordinates $[P_0XP_0,P_0YP_0]=\frac{1}{2iB}$. 
Our analysis here concerns the relative motion. This might seem 
problematic as the potential $V(|\vec{r}|)$ lifts the degeneracy of 
the Landau levels so that one can apparently no longer think of projection 
onto Landau levels, and particularly the lowest Landau level. Closer 
inspection of the argument in Ref.\cite{ouvry2} reveals, however, that 
the degeneracy is not essential. Indeed, the only requirement is that 
the subspace on which is to be projected is infinite dimensional as 
the non-commutative coordinates can only be realized in this case. 
A natural generalization of the analysis in Ref.\cite{ouvry2} would 
therefore be to identify a low energy infinite dimensional subspace on 
which to perform the projection. In the case of short range interactions 
$V(|\vec{r}|)$, for which the interaction energy scale is much less than 
the cyclotron frequency, which is the situation normally assumed, this 
can still be done. The reason is that the spectrum for the relative 
motion will clearly be close to that of the Landau problem for large 
values of the relative angular momentum as the particles are then well 
separated. For small values of the angular momentum the potential will 
have its main effect. However, if the interaction energy scale is much 
less than the cyclotron frequency, one will still have well separated 
bands of eigenstates with the cyclotron frequency the energy scale 
determining the separation between bands and the interaction energy scale 
the separation within bands. We can therefore identify an infinite 
dimensional low energy subspace as the lowest Landau level perturbed 
by the interaction and proceed to study the commutation relations of 
the relative coordinates projected onto this subspace.

This paper is organized in the following way. In section \ref{general} 
we describe the general projection procedure onto the low energy subspace. 
In section \ref{soluble} we apply this procedure to a number of exactly 
soluble interacting models to obtain insight into the underlying physics.  

\section{General projection on the low energy sector}
\label{general}
We start by recalling a few basic facts about the Landau problem. 
A particle moving on a plane, subjected to a perpendicular constant 
magnetic field $B$, has a discrete set of energy eigenstates, known as 
Landau levels, and are labelled as $|n, \ell\rangle$, where $n$ and 
$\ell$ are integers labelling the various Landau levels $(n)$ and 
the degenerate angular momentum eigenstates with integer eigenvalues 
$\ell(\geq-n)$ within the same Landau level $n$.

We focus on the relative motion of the two particles described by 
the second part of the Hamiltonian (\ref{ham2}),
\begin{equation}
\label{hamrel}
H =\left(\vec{p}-\frac{1}{2}\vec{A}(\vec{r})\right)^2+\tilde V(|\vec{r}|).
\end{equation}
From the rotational symmetry this problem can be solved as usual through 
the separation of variables and the wave functions have the generic form
\begin{equation}
\psi_{n,\ell}(\vec{r})=\langle \vec{r}|n,\ell\rangle=
R_{n,\ell}(r)e^{i\ell\phi},
\end{equation}
where $R_{n,\ell}$ solves the radial equation
\begin{equation}
\label{radial}
\left[-\frac{\partial^2}{\partial r^2}-\frac{1}{r}
\frac{\partial}{\partial r}+\frac{\ell^2}{r^2}-\omega_{c}\ell+
\frac{1}{4}\omega_{c}^{2}r^2 + V(|\vec{r}|)\right]R_{n,\ell}=
E_{n,\ell}R_{n,\ell},
\end{equation}
$n$ is the principle quantum number and $\omega_{c}=B/2$ is half of 
the cyclotron frequency. Under the conditions discussed in 
section \ref{intro} the separated bands of eigenstates will be labelled 
by the principle quantum number, $n$, while the states within a band will 
be labelled by the angular mometum $\ell$. In particular we assume that 
the lowest energy states are described by $n=n_0$ (say) and 
$\ell=0,1,2\ldots$ where we noted from (\ref{radial}) that a change in 
sign of the angular momentum will require an energy of the order of 
the cyclotron frequency, so that negative angular momenta will not occur 
in the low energy sector, {\it i.e.}, we are restricting to the lowest 
Landau level, perturbed by interactions. 
  
We can now construct the projection operator on the low energy sector as
\begin{equation}
\label{proj}
P_0=\sum_{\ell=0}^\infty |n_0,\ell\rangle\langle n_0\ell|.
\end{equation}
We now compute the projected relative coordinates
\begin{eqnarray}
\label{projop}
P_{0}xP_{0}&=&\sum_{l,l^{\prime}=0}^{\infty}
\langle n_0,\ell^{\prime}|x|n_0,\ell\rangle|n_0, \ell^{\prime}\rangle
\langle n_0, \ell|\,,\nonumber\\
P_{0}yP_{0}&=&\sum_{l,l^{\prime}=0}^{\infty}
\langle n_0,\ell^{\prime}|y|n_0,\ell\rangle|n_0, \ell^{\prime}\rangle
\langle n_0, \ell|\,,
\end{eqnarray}
with
\begin{eqnarray}
\label{me}
\langle n_0,\ell^{\prime}|x|n_0,\ell\rangle&=&\Omega_{\ell^\prime,\ell}
\left(\delta_{\ell^{\prime},\ell+1}+\delta_{\ell^{\prime},\ell-1}\right)
\,,\nonumber\\
\langle n_0,\ell^{\prime}|y|n_0,\ell\rangle&=&-i\Omega_{\ell^\prime,\ell}
\left(\delta_{\ell^{\prime},\ell+1}-\delta_{\ell^{\prime},\ell-1}\right)
\,,\nonumber\\
\Omega_{\ell^\prime,\ell}&=&\pi\int_0^\infty\; drr^2R^*_{n_0,\ell^{\prime}}
R_{n_0,\ell}\,.
\end{eqnarray}
The commutator of the relative coordinates then yields
\begin{equation}
\label{commrel}
\left[P_0xP_0,P_0yP_0\right]=2i\sum_{\ell=0}^{\ell=\infty}
|\Omega_{\ell,\ell+1}|^2\left[|n_0,\ell+1\rangle\langle n_0,\ell+1|-
|n_0,\ell\rangle\langle n_0,\ell|\right].
\end{equation}

We now simply have to compute the matrix elements $\Omega_{\ell^{\prime},\ell}$
to determine the commutator. For some potentials this can be done 
analytically and exactly, but in most cases one has to resort to 
approximations. In this regard we note that since the potential has 
radial symmetry, it will not mix different angular momentum sectors. 
Within a particular angular momentum sector there is of course no 
degeneracy of the Landau states, so that one can safely apply 
perturbation theory to compute the radial wave-functions $R_{n_0,\ell}$ 
and therefore matrix elements $\Omega_{\ell^{\prime},\ell}$. Indeed, 
this corresponds to a $1/B$ expansion.

When the interaction is switched off ($V(r)=0$) the radial wave-functions 
are those of the Landau problem and this result is easily seen 
to reduce to (\ref{commutator}), except for a factor of two. The reason 
for this is simply that since we are working with the relative coordinates 
between two particles this commutator should yield in the non-interacting 
case the minimal area occupied by two particles, which is consistent 
with (\ref{commutator}).

In contrast to the non-interacting case (\ref{commutator}) this commutator 
is in general no longer proportional to $P_0$. However, since we are 
dealing with a central potential, the different angular momentum 
($\ell$) sectors decouple and one can interpret this result as 
a non-commutative theory with an effective $\ell$ dependent non-commutative 
parameter in the same spirit as was done in Ref.\cite{scholtz}.  

As the area occupied by the two particles will increase with increasing 
relative angular momentum, one can deduce from (\ref{commrel}) and 
(\ref{uncertain}) an absolute lower bound to the average area that 
a particle in the low energy sector may occupy 
\begin{equation}
2\Delta A=4|\Delta (P_0xP_0)||\Delta (P_0yP_0)|\geq 4|\Omega_{0,1}|^2.
\end{equation}
The factor of two on the left is required as the right hand side is 
the average area occupied by two particles, as pointed out earlier.

\section{Non-commutativity in some soluble models}
\label{soluble}

In this section we study the non-commutative structure that arises in 
a number of soluble interacting models to gain deeper insight into 
the underlying physics.

\subsection{Harmonic oscillator}
\label{harosc}

\begin{figure}[t]
\begin{center}
	\epsfig{file=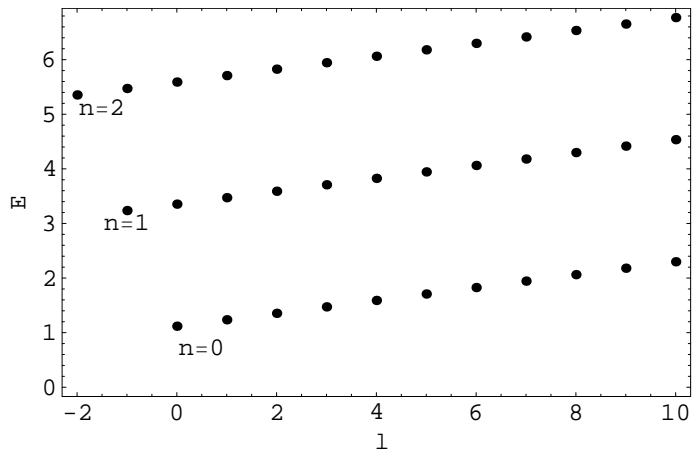,height=6.0cm,clip=,angle=0}
	\caption{The spectrum for the harmonic oscillator 
potential $\omega_c=1$ and $\lambda=0.5$.}
	\label{Fig1}
\end{center}
\end{figure}

We take for $V(|\vec{r}|)=\frac{\lambda^2}{4} r^2$.  This is not a 
short range potential and the spectrum will not approach that of the 
Landau problem for large values of $\ell$. Indeed, here one gets a 
spectrum linearly growing in $\ell$ (see Fig.\ref{Fig1}) so that 
one cannot claim that projection onto the lowest principle quantum number 
will correspond to the lowest energy sector. However, as was pointed out 
in the introduction, one can in principle project onto any 
infinite dimensional subspace, not necessarily just the lowest energy, 
and that is the spirit in which the current calculation is done.

In this case it is easy to solve the radial equation for the 
lowest principle quantum number:
\begin{eqnarray}
\label{harmonic}
R_{0,\ell}&=&N_{\ell}r^\ell
\exp(-\frac{1}{4}\sqrt{\omega_{c}^2+\lambda^2} r^2)\quad;\quad \ell\geq 0,\\
N_{\ell}&=&\frac{\left(\omega_{c}^2+\lambda^2\right)^{(\ell+1)/2}}
{\sqrt{\pi 2^{\ell+1}\Gamma(\ell+1)}}.
\end{eqnarray}
The spectrum is given by 
\begin{equation}
E_{0,\ell}=\ell(\sqrt{\omega_{c}^2+\lambda^2}-\omega_c)+
\sqrt{\omega_{c}^2+\lambda^2}
\end{equation}
and is linearly growing with increasing $\ell$. The spectrum and 
eigenfunctions for higher quantum numbers can of course also be solved 
easily and projection onto those subspaces can also be done. The full 
spectrum is given by $E_{n,\ell}=\sqrt{\omega_c^2+\lambda^2}(2n+1)+
\ell(\sqrt{\omega_c^2+\lambda^2}-\omega_c)$, $n=0,1,2\ldots$, $\ell\geq -n$ 
and is shown in Fig.\ref{Fig1} for $\omega_c=1$ and $\lambda=0.5$. 
As no new features appear we restrict ourselves here to the solutions 
with the lowest principle quantum number.

The commutator of the relative coordinates can now be evaluated 
from (\ref{commrel}) and yields
\begin{eqnarray}
\label{harcomm}
\left[P_0xP_0,P_0yP_0\right]&=&\frac{i}{\sqrt{\omega_{c}^2+\lambda^2}}
\sum_{\ell=0}^{\ell=\infty}(\ell+1)\left[|n_0,\ell+1\rangle\langle n_0,\ell+1|-
|n_0,\ell\rangle\langle n_0,\ell|\right]\nonumber\\
&=&\frac{1}{i\sqrt{\omega_{c}^2+\lambda^2}}P_0\nonumber\\
&=&\frac{2}{iB}(1+\frac{4\lambda^2}{B^2})^{-1/2}\,.
\end{eqnarray}
In the last line we have noticed that $P_0$ is just the identity on 
the projected subspace. As was discussed in general, we note that when 
the interaction is switched off ($\lambda=0$), this result differs 
by a factor of two from (\ref{commutator}).  

The first important point to note from this computation is that, 
generically, the non-commutative parameter is renormalized by 
the interactions.  
 
We can follow the same heuristic line of reasoning as for the free case 
to compute the filling fractions at which the interacting quantum Hall 
fluid behaves incompressibly. The filling factor is 
$\nu=\frac{N}{M}=\frac{2N}{AB}$. Arguing as in section \ref{general} 
it follows from (\ref{harcomm}) that the average area occupied by 
a particle is strictly bounded from below by 
$2\Delta A=4|\Delta (P_0xP_0)||\Delta (P_0yP_0)|\geq 4/B
\sqrt{1+\frac{4\lambda^2}{B^2}}\equiv 2\Delta A_0$. At maximum filling 
of the $p$ lowest Landau levels (bands) one expects the particles to occupy 
the minimum allowed area, {\it i.e.}, $N\Delta A_0=pA$ and 
$\nu=p\sqrt{1+\frac{4\lambda^2}{B^2}}$, $p$ integer. As the next available 
states are in the next Landau level, which are still separated on an energy 
scale of the cyclotron frequency under the assumption that 
the interaction energy scale is much less than the cyclotron frequency, 
one expects that the quantum fluid will be incompressible at these values 
of the filling. Note that these filling fractions are larger than 
the non-interacting values. This is easily understood from the attractive 
nature of the interaction which, effectively, enhances the magnetic field.

\subsection{Inverse square potential}
\label{inverse}

We take for $V(|\vec{r}|)=\frac{2\lambda^2}{r^2}$. This Hamiltonian is 
very similar in structure to the Hamiltonian of a charged particle moving 
in a plane and coupled to the gauge potential 
$A_{i}=-\frac{\alpha}{r^2}\epsilon_{ij}x^j$ corresponding to a singular 
flux tube located at the origin, augmented by a harmonic potential. 
We investigate this case in detail in the next section as it is of 
particular importance in quantum Hall systems. Taking a cue from 
the wave function of this Hamiltonian \cite{ouvry3},
the lowest energy wave functions ($n=0, \ell\geq0$) are obtained by 
making the following ansatz:
\begin{eqnarray}
\psi_{n=0, \ell}(r, \phi)= N_\ell r^{\Lambda(\ell)}e^{i\ell\phi}
\exp\left(-\frac{\omega_{c}}{4}r^2\right)
\label{19}
\end{eqnarray}
where $\Lambda(\ell)$ is some unknown quantity which will get fixed 
by (\ref{radial}). The solution for $\Lambda(\ell)$, the exact low energy 
eigenvalues $E_{n=0,\ell}$ and the normalisation constant $N(\ell)$ are 
given by,
\begin{eqnarray}
\Lambda(\ell) = \left(\ell^2+2\lambda^2\right)^{1/2}\;,
\label{20}
\end{eqnarray}
\begin{eqnarray}
E_{n=0, \ell}=\left[\left(\ell^2+2\lambda^2\right)^{1/2}-\ell+1\right]
\omega_c\;,
\label{21}
\end{eqnarray}
\begin{eqnarray}
N_\ell=\left[\frac{\omega_{c}^{\Lambda(\ell) +1}}{\pi 2^{\Lambda(\ell)+1}
\Gamma(\Lambda(\ell)+1)}\right]^{1/2}\;.
\label{21a}
\end{eqnarray}

\begin{figure}[t]
\begin{center}
	\epsfig{file=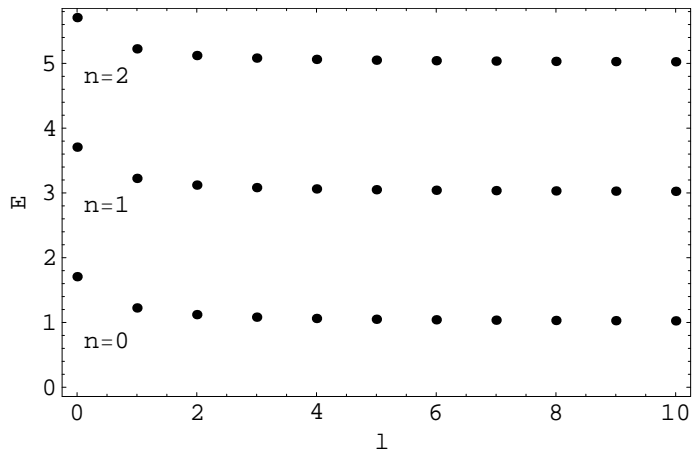,height=6.0cm,clip=,angle=0}
	\caption{The spectrum for the inverse square potential 
with $\omega_c=1$ and $\lambda=0.5$.}
	\label{Fig2}
\end{center}
\end{figure}

In this case the eigenvalues and eigenfunctions for higher Landau 
levels can also be solved as in Ref.\cite{ouvry3}. The full spectrum is 
given by $E_{n, \ell}=\left[2n+\left(\ell^2+2\lambda^2 \right)^{1/2}-
\ell+1\right]\omega_c$, $n=0,1,2\ldots$, $\ell\geq 0$ and is shown in
Fig.\ref{Fig2} for $\omega_c=1$ and $\lambda=0.5$. The expected features 
for a short range repulsive interaction can clearly be seen from this graph.

The commutator of the relative coordinates can now be evaluated 
from (\ref{commrel}) and yields
\begin{eqnarray}
\left[P_{0} xP_{0}, P_{0} yP_{0}\right]
=\frac{2i}{B}\sum_{\ell=0}^{\infty}F(\ell)^2
\left[|0, \ell+1\rangle\langle0, \ell+1|-
|0, \ell\rangle\langle0, \ell|\right]\;,
\label{23}
\end{eqnarray}
where $F(\ell)$ is given by
\begin{eqnarray}
F(\ell)= \frac{\Gamma\left(\frac{\Lambda(\ell)+\Lambda(\ell+1)+3}{2}\right)}
{[\Gamma(\Lambda(\ell)+1)\Gamma(\Lambda(\ell+1)+1)]^{1/2}}\;.
\label{24}
\end{eqnarray}

Note that in this case the right hand side of (\ref{23}) is not proportional 
to the projection operator $P_{0}$ and, as pointed out earlier, one 
should interpret this as an effective non-commutative theory with 
an $\ell$-dependent renormalized non-commutative parameter.

Note that contrary to what one might naively expect, the lower bound of 
the area of the particle in angular momentum sector $l$, given in terms 
of the quantities $|F(\ell-1)^2-F(\ell)^2|$, are not monotonically
increasing functions of $\ell$. To understand this, one must note that 
this lower bound is only achieved for minimum uncertainty states. 
The actual area is to be computed from $\langle r^2\rangle$ in the 
appropriate eigenstate, which is indeed a monotonically
increasing function of $\ell$. One therefore concludes that
the corresponding expression, evaluated at $\ell=0$ gives an absolute
lower bound. Arguing as before, it follows from (\ref{23}) that the 
average area occupied by a particle is strictly bounded from below 
by $2\Delta A=4|\Delta (P_0xP_0)||\Delta (P_0yP_0)|\geq \frac{4F(0)^2}{B}
\equiv 2\Delta A_0$. As before the filling fractions at which the fluid 
is incompressible are $\nu=\frac{p}{F(0)^2}$, $p$ integer. 
As $F(0)^2\geq 1$ this yields a fractional filling factor.  

\subsection{Singular magnetic fields}
\label{anyon}

In this section we consider the relative motion of the two particles 
without any interaction, but with a singular flux tube located at 
the position of the particles. To obtain the appropriate 
Hamiltonian \cite{ouvry3}, we perform a singular gauge transformation 
in the relative coordinate on the Hamiltonian (\ref{ham2}). To be precise 
we perform the gauge transformation $e^{i\alpha\phi} H e^{-i\alpha\phi}$ 
with $\phi=\tan^{-1}\left(\frac{y}{x}\right)$, with $y$ and $x$ 
the components of the relative coordinates. Dropping the center of mass 
part of (\ref{ham2}), which is not affected by the gauge transformation, 
the gauge transformed Hamiltonian for the relative coordinate, 
which corresponds to a singular flux tube inserted at the position of 
the particles, reads \cite{ouvry3}
\begin{equation}
\label{anyonham} 
H =\left(\vec{p}-\frac{1}{2}\vec{A}\right)^2\,,
\end{equation}
where
\begin{equation}
A_{i}=-(\frac{B}{2}+\frac{2\alpha}{r^2})\epsilon_{ij}x^{j},\; B\geq 0\;.
\end{equation} 
Written out explicitly the Hamiltonian reads
\begin{equation}
H=-\frac{\partial^2}{\partial r^2}-\frac{1}{r}\frac{\partial}{\partial r}+
\frac{1}{r^2}\left(i\frac{\partial}{\partial \phi}+\alpha\right)^2+
i\omega_{c}\frac{\partial}{\partial\phi}+\frac{1}{4}\omega_{c}^{2}r^2+
\alpha\omega_c\;.
\end{equation}
As in the composite fermion paradigm, we choose $\alpha\leq 0$ so that 
it leads to an effective reduction of the magnetic flux seen by 
the particles. The low energy eigenfunctions and spectrum are easily 
found to be \cite{ouvry3}
\begin{eqnarray}
\psi_{0,\ell}&=&N_\ell r^{|\ell-\alpha|}e^{i\ell\phi}
e^{-\frac{\omega_c r^2}{4}},\;\ell\geq 0\,,\nonumber\\
N_\ell&=&\left[\frac{\omega_c^{|\ell-\alpha|+1}}
{\pi 2^{|\ell-\alpha|+1}\Gamma\left(|\ell-\alpha|+1\right)}\right]^{1/2}\,,\\
E_{0,\ell}&=&\omega_c\,.\nonumber
\end{eqnarray}

The commutator of the relative coordinates yields from (\ref{commrel})
\begin{eqnarray}
\label{sing}
\left[P_{0} xP_{0}, P_{0} yP_{0}\right]
&=&\frac{2i}{B}\sum_{\ell=0}^{\infty}\left(l+|\alpha|+1\right)
\left[|0, \ell+1\rangle\langle0, \ell+1|-|0, \ell\rangle\langle0, \ell|\right]
\nonumber\\
&=&\frac{2}{iB}\sum_{\ell=0}^{\infty}\left(|\alpha|\delta_{0,\ell}+1\right)
|0, \ell\rangle\langle 0, \ell|.
\end{eqnarray}
 
As before, it follows from (\ref{sing}) that the average area occupied 
by a particle is strictly bounded from below by $2\Delta A=
4|\Delta (P_0xP_0)||\Delta (P_0yP_0)|\geq \frac{4(1+|\alpha|)}{B}\equiv 
2\Delta A_0$ and the filling fractions at which the fluid is incompressible 
are $\nu=\frac{p}{1+|\alpha|}$, $p$ integer. Keeping in mind the phase 
factor associated with the singular gauge transformation, unchanged 
statistics requires, as usual, that one must choose $\alpha=2k$ with 
$k$ a negative integer. This choice indeed yields the fractional fillings as obtained 
from the composite fermion picture \cite{jain} when appropriate 
choices of $p$ and $k$ are made.

\section{Discussion and conclusions}

We discussed the role that interactions play in the non-commutative 
structure that arises when the relative coordinates of two interacting 
particles are projected onto the lowest Landau level. It was shown that 
the interactions in general renormalize the non-commutative parameter 
away from the non-interacting value $\frac{1}{B}$. The effective 
non-commutative parameter is in general also angular momentum dependent, 
as was also found from other considerations in Ref.\cite{scholtz}.
An heuristic argument, based on the non-commutative coordinates, was 
given to find the filling fractions at incompressibility and the results 
are consistent with known results in the case of singular magnetic fields.
It should, however, be kept in mind that this argument was very simplistic 
as all possible many-body correlations were ignored. Probably due to 
this oversimplification this argument cannot explain, for a general 
short range repulsive interaction, the quantized values of the filling 
fraction at incompressibility observed in the fractional quantum Hall 
effect. Indeed, from naive perturbative considerations in the above 
setting one would expect that the (screened) Coulomb interaction will 
have only a perturbative effect on the non-commutative parameter and 
filling fraction, which is certainly not the case. As in other treatments, 
it is only when one already assumes the existence of composite fermions, 
as was done in section \ref{anyon}, that the quantized filling fraction 
can be explained. The apparently non-perturbative microscopic origin
of composite fermions as effective non-interacting degrees of freedom to describe 
the Coulomb interacting quantum Hall fluid is, indeed, still an illusive 
and controversial issue \cite{dyakonov}.           

\section*{Acknowledgements}

This work was supported by a grant under the Indo--South African research 
agreement between the Department of Science and Technology, Government of 
India and the South African National Research Foundation. FGS would like 
to thank the S.N. Bose National Center for Basic Sciences and the 
Institute of Nuclear Physics, University of Louvain, for their hospitality 
in the periods that parts of this work were completed. BC would like to 
thank the Institute of Theoretical Physics, Stellenbosch University for 
their hospitality during the period when part of this work was completed.

\end{document}